\documentclass[12pt]{iopart}

\usepackage{bm}
\usepackage{graphics}
\usepackage[pdftex]{graphicx}
\usepackage{epstopdf}
\begin{document}

\title{The Second Law of Thermodynamics and Entropy-Decreasing Processes With $^4$He Superflows }

\author{Yongle Yu}

\address{State Key Laboratory of Magnetic Resonance and Atomic and \\
Molecular Physics,
  Wuhan Institute of Physics and Mathematics, \\Chinese Academy of Science,
   West No. 30 Xiao Hong Shan, Wuchang, \\ Wuhan, 430071, China}
\ead{yongle.yu@wipm.ac.cn}

\begin{abstract} 
We review on a recently proposed quantum exception to the second law
of thermodynamics. We emphasize that  $^4$He superflows, like any other
forms of flows, shall carry entropy or heat in a thermal environment.
Following that, one can use a heterogeneous 
 $^4$He superflow loop to realize entropy-decreasing processes. 
We also mention that the heat content of a superflow has an unusual
dependence on flow velocity, which is an important factor contributing
to the entropy-decreasing processes.

\end{abstract}
\vspace{2pc}



\vspace{2pc}


The second law of thermodynamics (SLT)
 prohibits the decrease
of the entropy in an isolated system. It rules out
the desired possibility of  utilizing the endless colossal 
 thermal energy 
in the environment 
as the  major energy source for the civilization. But is
this SLT truely universal?  We know that first law of
 thermodynamics is exact universal, because it is 
 a manifestation of conservation of energy and 
quantum mechanics obeys the conservation of energy as a built-in
law. Unlike the first law, SLT doesn't corresponds 
to a built-in rule in quantum mechanics. Imagine a situation in 
which that numberless physical processes obey SLT, 
but there is one quantum process which violates SLT, then in this situation
 we shall accept that SLT is not
an exact law of nature and that there is an exception to SLT. 
It is quantum mechanics, rather than SLT, which is the true
governing law of nature. In \cite{slte}, we show that some $^4$He-superflow-involving
quantum 
processes contradict SLT, and
that it is possible to  convert the thermal energy
in the environment into useful energy. We shall review this
quantum exception to  SLT.

The entropy-decreasing quantum processes
rely on a fundamental property of  $^4$He superflows.
Whether does 
 a $^4$He superflow carry entropy or thermal energy?
The phenomenological two-fluid model of superfluid $^{4}$He 
 gives a no answer. But this is at odds with
quantum mechanics. A $^4$He superflow corresponds to 
a number  ($N$) of $^4$He atoms flowing together. If this superflow carries
zero entropy and heat, that is equivalently to say that
the quantum state of the $N$ $^4$He atoms in the superflow
is the lowest energy state at given momentum (or at given current
value). We know that the time evolution of the whole system
(the superflow
plus its environment) is governed by the many-body Schrodinger
equation. In  this time evolution, the quantum state of the environment is
a mixed quantum state, corresponding to a certain temperature. The subsystem
of the $N$ $^4$He atoms (superflow) can be expected to be a mixed
quantum state too. It is hard to imagine how this  subsystem  could  reach
 a zero entropy and zero thermal energy
quantum state following the time evolution. Even if at a moment
this superflow reaches this zero entropy 
quantum state dramatically, it shall evolve into a mixed state at
the following moments. In other words, the superflow shall be a 
thermal flow in a thermal environment, similar to any conventional
flows. 
\begin{figure}%
\includegraphics[scale=0.7]{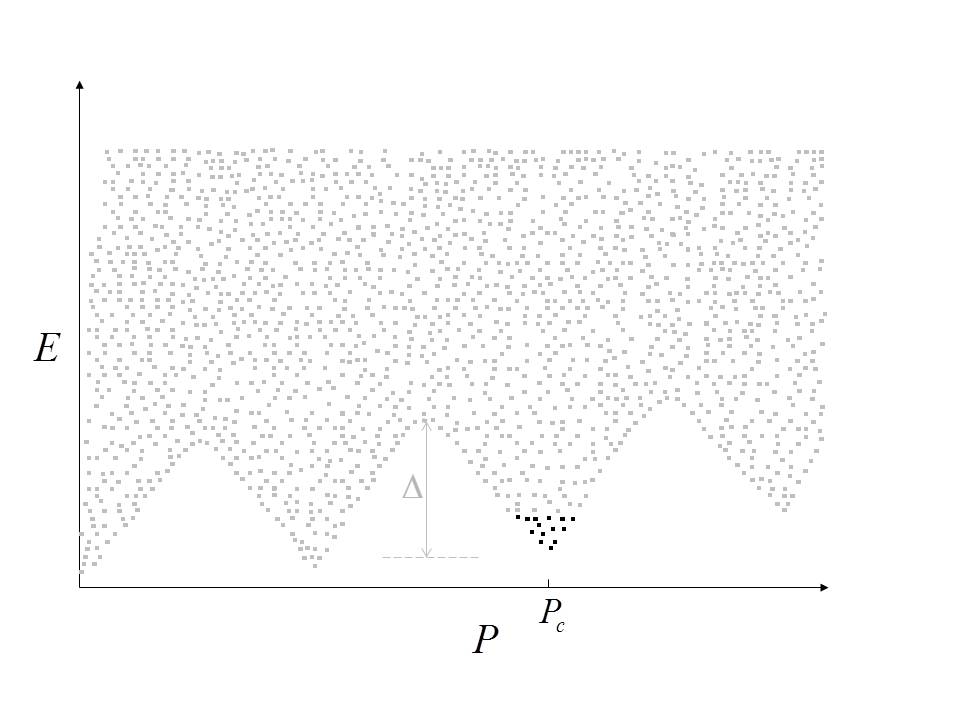}
\caption{A schematic plot of many-body eigen levels of a superfluid in the $E-P$ Plane. 
Dots (in gray and in black)
 represent levels. The horizontal and vertical coordinates of a dot correspond to
the momentum and the energy of the level, respectively. $\Delta$ denotes the height of
an energy barrier between two "valleys". Dots in black are the occupied levels of
the system (at low temperature), corresponding to a superflow state with a momentum of $P_c$.}
\label{fig:sflevels}
\end{figure}

Can we understand that a $^4$He  superflow is both thermal 
and superfluidic ( i.e., keeping its current from decay)?
 A recently developed 
quantum theory of superfluidity \cite{yu, bloch, leggett} 
provides an affirmative answer.
  This microscopic theory shows that
the many-body spectrum of a superfluid has a characteristic feature 
\cite{yu, bloch, leggett}. Consider  a superfluid   composed of  $N$ particles 
in a limited geometry, 
the Hamiltonian of this many-body system can be written as
\begin{equation}
   \widehat H= - \sum_{i=1}^{N} \frac{\hbar^2}{2M}
    \frac{\partial^2} {\partial 
  \bm{r}_i^2}+  \sum_{i<j}^N V(\bm{r}_i-\bm{r}_j),
\end{equation}
where $M$ is the mass of a particle and $V(\bm{r})$ represents the inter-particle interaction. 
Consider
the eigenspectrum of the Hamiltonian operator, labelled  by $\kappa $,
\begin{equation}
\widehat H\psi_\kappa(\bm{r}_1, \bm{r}_2, ..., \bm{r}_N)= E_\kappa \psi_\kappa (\bm{r}_1, \bm{r}_2, ..., \bm{r}_N),
\end{equation}
where $E_\kappa $ is eigenlevel and $ \psi_\kappa $ is the  eigenwavefunction. We shall
 consider also the  momentum carried by eigenwavefunction $\psi_\kappa $,
\begin{equation}
    P_\kappa = \langle \psi_\kappa (\bm{r}_1, \bm{r}_2, ..., \bm{r}_N)| \widehat{P} |
\psi_ \kappa (\bm{r}_1, \bm{r}_2, ..., \bm{r}_N)\rangle,
\end{equation}
where  $\widehat{P}$ is the total momentum operator along a superflow direction.
$\psi_\kappa$ is not required to be an eigenwavefunction to the operator $\widehat{P}$.

 If one schematically
plots the energy levels of this system at the energy-momentum ($E-P$)
plane (see Fig. \ref{fig:sflevels}), one then sees that the low-lying 
levels  form valley-like structures at
the boundary region \cite{yu, bloch, leggett}.  The relevant quantum states at low temperature 
corresponds to those levels within the bottoms of the valleys and one
can ignore the levels at higher energies. 
 A superflow state (at low temperature) corresponds
to some occupied  levels at a valley at a non-zero momentum. The system 
is not allowed to jump 
from the occupied levels at this valley to  left-side valleys
with less momenta, due to the energy barriers which separate
the valleys. The prohibition of inter-valley jumps ensures that a superflow 
 keeps its momentum from decay. For a given superflow, the occupation probability distribution 
at a relevant valley follows a Boltzmann distribution, namely,  the occupation
probability of a level at the valley  is proportional to $e^{-E/kT}$, 
where $E$ is the energy of
the level, $k$ is the Boltzmann constant and $T$ is the temperature of the superflow. 
Microscopically this thermal distribution is  caused by the quanta exchanges
between the superflow and its surroundings, which in turn is caused by 
microscopic interactions
between the atoms in the superflows and the particles (atoms or molecules) in the
surroundings. 

\begin{figure}%
\begin{center}
\includegraphics[scale=0.6]{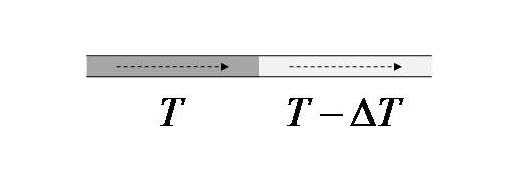}
\end{center}
\caption{The temperature of a $^4$He superflow changes when exiting a medium
to a different medium. }
\label{fig:sfPeltier}
\end{figure}

Once realizing that $^4$He  superflows are still thermal flows, one can expect
a phenomenon of  $^4$He  superflows  similar to the Peltier effect of electric currents.
When a $^4$He superflow passes from a medium to a different medium, it shall experience
a temperature change.  The heat content ( specific enthalpy) of a superflow in the first
medium as a function of temperature differs from the heat content function 
in the second medium.  For the conservation of energy, the temperature of 
the superflow shall be
adjusted  so that heat content remains the same (roughly) when the superflow
exits the first medium (see Fig. \ref{fig:sfPeltier}). In literature,
 the temperature change 
of $^4$He  superflow exiting a medium to a rather large empty space was
reported  \cite{mechcal} shortly after the discovery of superfluidity \cite{sf}.

One could note that the flow velocity of the superflow generally changes when
it passes from the first medium to the second medium. If the first medium has 
a porosity smaller than the second medium, then the flow velocity in the first
medium is larger than the flow velocity at the second medium. It can be shown that
the heat content of a superflow in a medium depends on the flow velocity \cite{yuVel},
 which is quite unusual from a classical viewpoint. Qualitatively, the larger
 the flow velocity,
the smaller the heat content. Such a dependence of a superflow's heat content
on the flow velocity could make a large contribution 
to this "Peltier effect" of
the superflow. 

\begin{figure}%
\begin{center}
\includegraphics[scale=0.6]{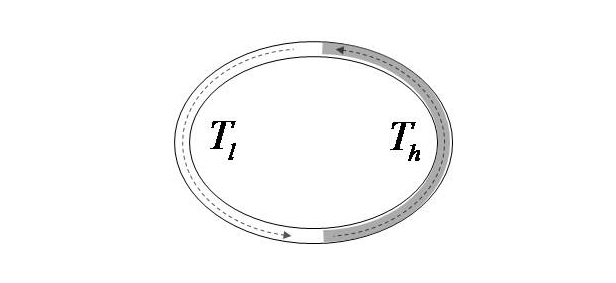}
\end{center}
\caption{A heterogeneous superflow loop. The temperature of the 
superflow in one medium differs from the the temperature of the
superflow in the other medium. This temperature difference is induced
by the superflow.
} 
\label{fig:HSloop}
\end{figure}

\begin{figure}%
\begin{center}
\includegraphics[scale=0.6]{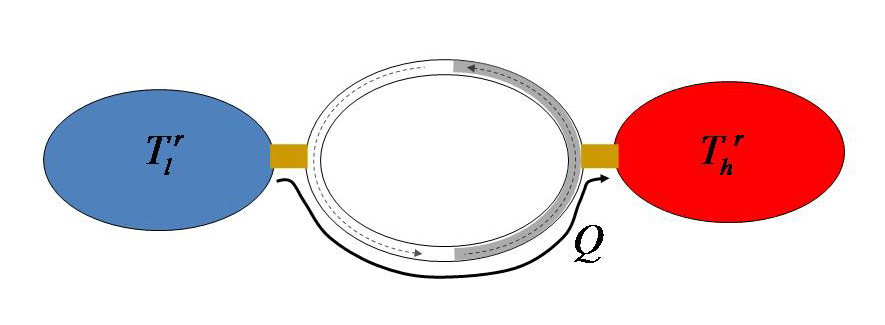}
\end{center}
\caption{An exception to SLT. There is a heat flow from the heat reservoir
at  $T{^r}_l$ to  another  heat reservoir at a higher temperature. 
} 
\label{fig:SLTCE2}
\end{figure}

In \cite{slte}, we consider a system referred as to a heterogeneous superflow loop (HSL). It is 
a circling ${^4}$He superflow filling a torus-shaped vessel, where half of 
the vessel is packed with one kind of medium and the other half is packed with a different
 medium \cite{medium} (see Fig. \ref{fig:HSloop}).  one can find an interesting stable temperature 
configuration along an isolated HSL. The superflow in one medium has a temperature
$T_h$ higher than the temperature of the superflow at the other medium (denoted by
$T_l$). With such a temperature difference, the heat content of superflow in
 one medium is (approximately) the same as the heat content of superflow in the other medium,
so that net heat transfer from one medium to the other, caused by the superflow, becomes
zero (approximately). To show a dramatical counterexample to SLT, 
 consider to prepare   two infinitely large heat baths 
(besides a HSL), with one bath at a temperature  $T^r_h$ and the other at a temperature
$T^r_l$.  $T^r_h$  and $T^r_l$ is set to satisfy  $ T_h > T^r_h>T^r_l>T_l$. Consider 
 making a good thermal contact between the high
 temperature part of the HSL and the bath at $T^r_h$, 
and making a good thermal contact between the low temperature part of HSL and the 
 bath at $T^r_l$.
Then, the superflow absorbs some heat from the heat reservoir at $T{^r}$, 
and this (moving) superflow transfers the heat to the high temperature part of the HSL,
and eventually the heat is passed to
  the heat reservoir at $T{^h}$ (see Fig. \ref{fig:SLTCE2}). 
This is a process in which the entropy of the whole system 
decreases. The entropy-decreasing process can run for a 
cosmologically long time, since that the decay of the superflow is prevented by 
some energy barriers \cite{reppy}. 
 
In \cite{slte}, we sketch out a prototype of a system 
 for extracting thermal energy
 from the environment. This potential energy approach
is environmentally friendly to a great extent. The energy-generating process
 can be purely physical, without consuming any materials or producing any
 unwanted materials.  It can be roughly estimated that a system with
one kilogram of liquid helium generates ideally
 a power  of several kilowatts or more,
 but some relevant cryogenic technology
needs to be developed for achieving such an efficiency.

To summarize,  a quantum exception to SLT can be constructed with  $^4$He superflows.

Financial
 supports from Chinese NSF (Grant No. 11474313), from CAS (Grant No. XDB21030300), from NKRDP  of China 
( Grant No. 2016YFA0301503), are acknowledged.


\begin{thebibliography}{10}
\bibitem{medium} Here a medium could also refer to vacuum, and to a specific geometric arrangement
of a superflow.
\end{thebibliography}
\end{document}